\documentclass[aps,prr,twocolumn,floatfix,superscriptaddress,showpacs]{revtex4}

\usepackage{amssymb,amsmath,amstext}                %%   American Physical Society math etc extensions
\usepackage{graphicx}                                               %%   Include figure files                                            %%   Manage links
\usepackage{epstopdf}                                               %%   help with eps -> pdf 
\usepackage{color}                                                     %%   change color
\usepackage{bm}                                                        %%   bold math                                    
\usepackage{appendix}                                              %%   formating appendicies
\usepackage[utf8]{inputenc}
\usepackage{bbold}
\usepackage{bbm}   
\usepackage{ulem}
\usepackage{latexsym}
\usepackage[colorlinks=true,citecolor=blue,linkcolor=magenta]{hyperref}
\usepackage{mathtools}
\usepackage{mathtools}

\def\be{\begin{equation}}
\def\ee{\end{equation}}
\def\bea{\begin{eqnarray}}
\def\eea{\end{eqnarray}}
\def\ra{\rangle}
\def\la{\langle}
\def\bi{\begin{itemize}}
\def\ei{\end{itemize}}
\def\ben{\begin{enumerate}}
\def\een{\end{enumerate}}

\definecolor{dgreen} {RGB}{0,100,0}
\newcommand{\revision}[1]{{\color{magenta}{{#1}}}}

\begin{document} 

\title{Can a bright soliton model reveal a genuine time crystal for a finite number of bosons?}

\author{Andrzej Syrwid} 
\email{syrwid@kth.se}
\affiliation{Department of Physics, KTH Royal Institute of Technology, Stockholm SE-10691, Sweden}
\affiliation{Instytut Fizyki Teoretycznej, 
Uniwersytet Jagiello\'nski, ulica Profesora Stanis\l{}awa \L{}ojasiewicza 11, PL-30-348 Krak\'ow, Poland}
\author{Arkadiusz Kosior} 
\affiliation{Institute for Theoretical Physics, University of Innsbruck, 6020 Innsbruck, Austria}
\affiliation{Instytut Fizyki Teoretycznej, 
Uniwersytet Jagiello\'nski, ulica Profesora Stanis\l{}awa \L{}ojasiewicza 11, PL-30-348 Krak\'ow, Poland}
\author{Krzysztof Sacha} 
\affiliation{Instytut Fizyki Teoretycznej, 
Uniwersytet Jagiello\'nski, ulica Profesora Stanis\l{}awa \L{}ojasiewicza 11, PL-30-348 Krak\'ow, Poland}

\begin{abstract}
We analyze time crystal effects in a finite system of bosons which form a bright soliton clump on the Aharonov-Bohm ring. In the large particle number limit, $N\rightarrow\infty$, this setup corresponds to the Wilczek model, where it  is known that the time crystal behavior cannot be observed in the ground state of the system because a spontaneously formed soliton does not move. Here, we show that while the spontaneous formation of a moving soliton in the ground state can occur for $N<\infty$, the soliton decays before it makes a single revolution along the ring and the time crystal dynamics is impossible. 
\end{abstract}


\date{\today}

\maketitle

%%%%%%%%%%%%%%%%%%%%%%%%%%%%%%%%%%%%%%%%%%%%%%%%%%%%%%%%%%%%%%%%%%%%%%%%%%%%%%%%%%%%%%%
% 

A genuine quantum time crystal would be a system revealing periodic evolution in its lowest energy state. Quantum time crystals were proposed by Wilczek who considered attractively interacting bosons on a one-dimensional ring \cite{Wilczek2012}. In the presence of a magnetic-like flux $\alpha$ penetrating the ring (the so-called Aharonov-Bohm ring), bosons were expected to form a bright soliton wavepacket that would move  periodically along the ring even if the energy of the system was minimal. It turned out that it was not possible because in the limit of the number of bosons $N\rightarrow\infty$, the bright soliton that forms spontaneously in the lowest energy solution was always stationary \cite{Bruno2013,Syrwid2017}. Generally, for many-body systems with two-body interactions a genuine quantum time crystals cannot be formed in the thermodynamic limit \cite{Bruno2013b,Watanabe2015,Watanabe2020} (however, note that for multi-body interactions, a time crystal behavior in a system's ground state is possible \cite{Kozin2019}). For reviews see \cite{Sacha2017rev,khemani2019brief,Guo2020,SachaTC2020}.

While it is clear that in the limit of $N\rightarrow\infty$ a genuine time crystal in a soliton model proposed by Wilczek cannot  exist, one may ask a question 
whether a glimpse of a time crystal, i.e. at least a single oscillation of a solitonic clump in the ground state, could be observed in a finite particle system (see Ref.~\cite{Huang2017a} and also a discussion in Refs.~\cite{Ohberg2019,SyrwidKosiorSacha2020,ReplyOhbergWright2020,sksPRR2020,OWcomment2020}). In other words, is it possible that for large but finite $N$ the lifetime of a spontaneously formed moving soliton could be at least longer than the period of its evolution along the ring? This question is addressed in the present paper.

We consider $N$ bosons on a ring of the unit circumference which interact via contact potential in the presence of a magnetic-like flux $\alpha$. The corresponding many-body Hamiltonian reads
\be
H=\sum_{i=1}^N\frac{(p_i-\alpha)^2}{2}+\frac{g_0}{2}\sum_{i\ne j}^N\delta(x_i-x_j),
\ee
where $g_0<0$ determines the strength of the attractive interactions and we assume $m=\hbar=1$. For $\alpha=0$ and $g_0(N-1)<-\pi^2$, it is known that in the mean-field description the ground state of the system breaks spontaneously the space translation symmetry and a bright soliton is formed \cite{Syrwid_2021,footnote1}. In the case of a single particle on the ring, the momentum is quantized, i.e. $p_n=2\pi n$, and a proper choice of $\alpha$ leads to a non-vanishing probability flux along the ring. Indeed, if $\alpha$ is not an integer multiple of $2\pi$, then the particle velocity $p_n-\alpha\ne 0$. Nevertheless, in such a case the corresponding particle density is spatially uniform due to translational invariance of the system. When we switch to the many-body case and a bright soliton forms spontaneously in the system ground state, the soliton velocity can be calculated by introducing the the center of mass position and the conjugate momentum which is the total momentum $P$ of all particles. 
The center of mass energy reads 
\be\label{Hamiltonian_CM}
H_N=\frac{(P-\alpha N)^2}{2N},
\ee
 and consequently the corresponding center of mass 
velocity can be determined as follows
\be
\frac{d H_N}{d P_n}=2\pi \frac{n}{N}-\alpha,
\label{CMvelocity}
\ee
where we have substituted a quantized value of the total momentum $P_n=2\pi n$. It is clear from Eq.~(\ref{CMvelocity}) that in the limit of $N\rightarrow\infty$ it is always possible to find such $n$ that the velocity vanishes and the spontaneously formed bright soliton does not move in the ground state of the system. However, for $N<\infty$, one can always choose $\alpha$ which leads to a non-zero center of mass velocity and thus  resulting in a motion of a soliton along the ring. If it is so, there is a question if the lifetime of the soliton is long enough to complete at least one full revolution along the ring before it decays due to quantum many-body effects.

%%%%%%%%%%%%%%            
\begin{figure}
\includegraphics[width=1\columnwidth]{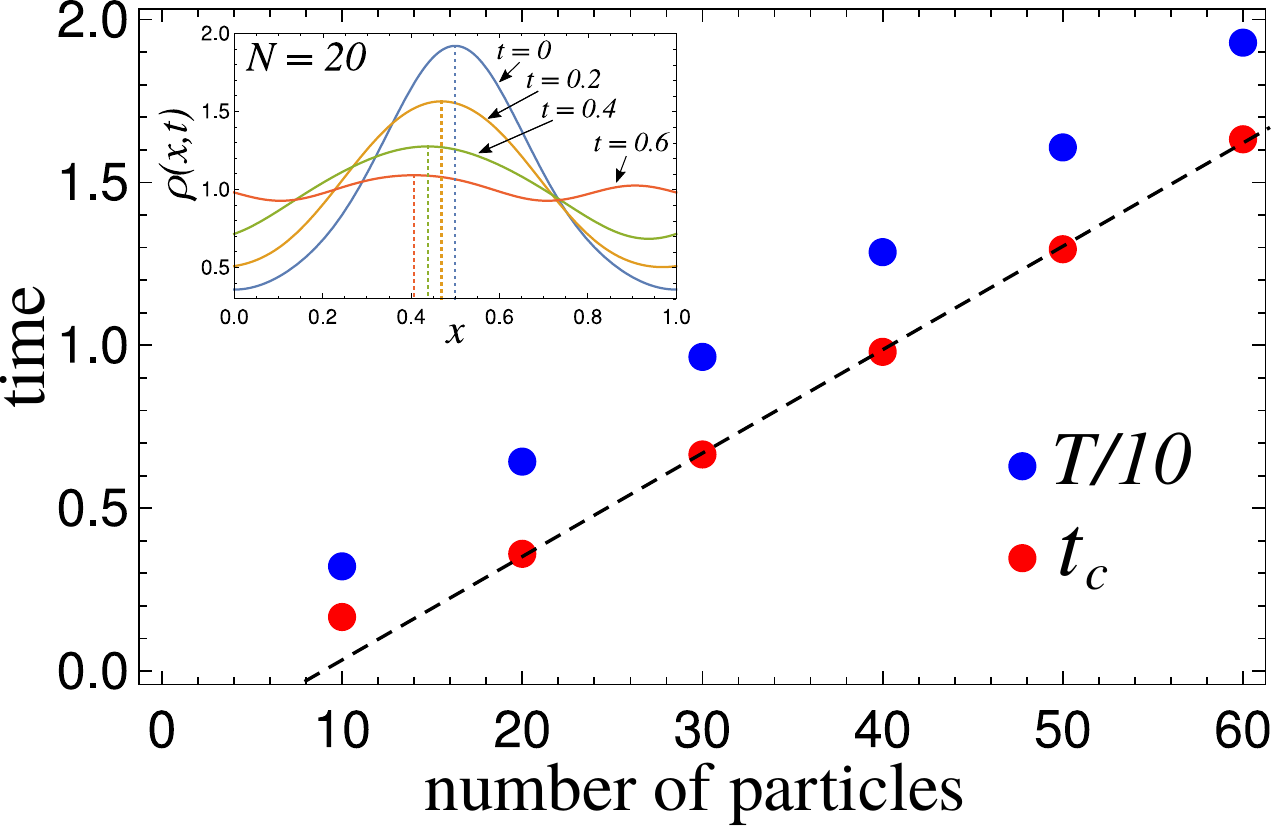} 
\caption{
Comparison between the lifetime $t_c$ of the soliton and the period $T=N/0.99\pi$ of the soliton's motion around the ring for different numbers of particles $N$. Note that $t_c$ (red dots) increases linearly with $N$ \revision{where $t_c\approx 0.032 N-0.28$ for $N\ge 20$} (dashed black line represents the linear fit) and is more than 10 times shorter than the period $T$ (note that blue dots correspond to $T/10$). For illustration, in the inset we present how the soliton structure visible in $\rho(x,t)$ dies out in time for $N=20$. Vertical dotted lines guide the eye and indicate the position of the soliton clump in four different time moments $t=\{0,0.2,0.4,0.6\}$. 
  }
\label{fig_1}   
\end{figure} 
%%%%%%%%%%%%%% 

To address this question it is sufficient to consider $\alpha N  \in(-\pi,\pi)$, for which the ground state corresponds to $n=0$. Note that the soliton velocity is maximized when the magnetic-like flux $|\alpha| \rightarrow \pi/N$. Thus, following the idea presented in Ref.~\cite{Syrwid2017}, we analyze a density-density correlation function
\begin{align}
\rho_2(x_2,t_2;x_1,t_1)&
\\ \nonumber
&\hspace{-0.5cm}\propto \la \hat\psi^\dagger(x_1,t_1) \hat\psi^\dagger(x_2,t_2)\hat\psi(x_2,t_2)\hat\psi(x_1,t_1)\ra,
\end{align}
for $\alpha N = 0.99\pi$ and the strength of the attractive interactions between particles given by $g_0(N-1)=-12.5$, where $\hat\psi$ is the bosonic field operator.
 Such a choice guarantees a formation of a bright soliton wavepacket that propagates with 99\% of the maximal velocity if we start with the ground state of the system consisting of $N<\infty$ bosons. To break the space translation symmetry possessed by the ground state of the translationally invariant system we perform an initial measurement of a single boson at $x_1=0.5$ and at $t_1=0$. Thanks to this, we can monitor a temporal behaviour of the single particle density after the initial measurement, $\rho(x,t)=\rho_2(x,t;0.5,0)$, which reveals a soliton-like wavepacket propagating along the ring with the expected velocity $-0.99\pi/N$ and decaying in time due to quantum many-body effects. The lifetime $t_c$ of the soliton can be estimated basing on the contrast quantity 
 \be\label{contrast}
 \mathcal{C}(t)=\frac{\mathrm{max}_x[\rho(x,t)] - \mathrm{min}_x[\rho(x,t)]}{\mathrm{max}_x[\rho(x,t)] + \mathrm{min}_x[\rho(x,t)]},
\ee 
see Ref.~\cite{Syrwid2017}. Here we define $t_c$  as the minimal time for which $\mathcal{C}(t_c)\approx 0.5\mathcal{C}(0)\approx 0.34$.  

In Fig.~\ref{fig_1} we show that in the soliton model regardless of how small $N$ one chooses, it is not possible to observe even a single revolution of a soliton before it spreads along the ring.

One could argue that after performing just  a single particle position measurement on the ground state the symmetry broken many-body state is far from being the perfect condensate where all of the particles occupy the mean-field solution. Indeed, as shown in the inset of Fig.~\ref{fig_1}, the symmetry broken state at $t=0$ reveals a soliton-like clump of width  
larger than the width of the corresponding mean-field soliton density. The reason for this is that after the measurement of a single particle the center of mass of a resulting soliton is delocalized on a distance comparable to the width $\ell_0\approx  2 \sigma_0$ of the mean-field soliton density~\cite{Sacha2017rev}, where according to the sech approximation, the standard deviation of the mean-field soliton density is given by $ \sigma_0=\pi/[\sqrt{3}|g_0|(N-1)]$. As shown in Ref.~\cite{Syrwid2017}, the condensate fraction will only approach unity after the measurement of a significant fraction of particles or with increasing the initial number of  particles $N$. The later, however, decreases the maximal speed of revolution of the ground state soliton and in the thermodynamic limit the soliton is stationary. On the other hand, increasing the number of measurements on the ground state at $t=0$ will lead to the stronger localization of the center of mass position, and consequently to the smaller width of the solitonic clump. This, in turn, decreases the lifetime of the soliton, which can be understood through a very straightforward argument. Namely, a strong localization of the center of mass in the position space is equivalent to a wide distribution in the momentum space  which leads to the faster spreading and to the faster decay of the soliton.

The above reasoning can be formulated more quantitatively.  We can assume that the wavefunction of the center of mass after an initial measurement of a single particle is approximated by a gaussian wavepacket of width $\ell_0$. Then, due to the time evolution by the Hamiltonian in Eq.~\eqref{Hamiltonian_CM}, its width expands as 
$\ell_{\mathrm CM}(t)  \propto  t/(N \ell_0)$ after the initial short time sublinear growth. 
Therefore,  the decay time $t_c$ scales as $t_c \propto N$.  We confirm this prediction in Fig.~\ref{fig_1}.  
On the other hand,  additional measurements of consecutive particles' positions drive the subsystem of the remaining particles the closer to the condensate product state, the more particles were measured. In the ideal situation, after the detection of a significant fraction of particles at $t=0$, the symmetry broken state describes particles occupying the same mean-field soliton solution. In such a case, according to the central limit theorem, the probability density of the center of mass position is given by a gaussian with a standard deviation $\sigma \approx \sigma_0/\sqrt{N}$, which leads to
 $\ell_{\mathrm CM}(t) \propto t/(\ell_0 \sqrt{N}) $,  with the decay time $t_c\propto \sqrt{N}$ \cite{Sacha2017rev}. 
 In conclusion, the measurement of many particles only deteriorates the decay time. Therefore, in this Letter we consider only the best case scenario where the symmetry of the ground state is broken by a single particle position measurement only. 
Note that a single particle measurement we perform is consistent with a  standard symmetry breaking definition in thermodynamic limit, where the symmetry breaking is equivalent to a periodic behavior of the second order correlation function.

In this Letter we have shown that a genuine time crystal behavior cannot be observed in a bright soliton model for a finite number of bosons $N$. Although the numerical simulations have been performed for a single choice of the interaction strength, $g_0(N-1)=-12.5$, it is a representative example of a general behavior. Since the speed of the soliton cannot be greater than $|v_{\mathrm{max}}|=\pi/N$, its minimal period of the revolution around the unit ring would be equal to $T_{\mathrm{min}}= N/\pi$. Nevertheless, the following reasoning shows that the lifetime of the soliton $t_c$ is also bounded from above, and cannot exceed $T_{\mathrm{min}}$. 
 
After a single position measurement at $t=0$ the width of the particle distribution in the symmetry broken state $\ell_{\mathrm{SB}}$  can be approximated as 
\be\label{sigma_sb}
\ell_{\mathrm{SB}}(t) \approx  \sqrt{\ell_0^2+\ell_{\mathrm CM}^2(t)}
,
\ee
where $\ell_0$ is the width of the corresponding mean-field soliton density and $\ell_{\mathrm CM}(t)$ denotes a width (double standard deviation) of the gaussian distribution describing the delocalization of the center of mass position. The latter quantity, $\ell_{\mathrm CM}(t)$, increases in time according to the solution of the standard quantum mechanics textbook problem
\be
\ell_{\mathrm CM}(t) \approx 2\, \sqrt{\frac{\ell_0^2}{4} +\left(\frac{2 t}{N\ell_0}\right)^2},
\ee
with the abovementioned assumption $\ell_{\mathrm CM}(t=0)\approx\ell_0$. Note that for $g_0(N-1) =-12.5$, in the sech approximation we have $\ell_0 \approx  0.3 $ and $\ell_{\mathrm{SB}}(t=0) \approx 0.4$, which is consistent with our numerical results, cf. the inset of Fig.~\ref{fig_1}.   
Let us for simplicity assume that the soliton can be recognized in the particle distribution if $\ell_{\mathrm{SB}}$  does not exceed the length of the ring, i.e., in our units, if $\ell_{\mathrm{SB}} \leq 1$. 
  Solving Eq.~(\ref{sigma_sb}) with respect to $t$, we get
 \be
 t(\ell_{\mathrm{SB}})\approx \frac{N}{4}  \ell_0 \sqrt{\ell_{\mathrm{SB}}^2-2\ell_0^2}.
 \ee
 This simple analysis allows us to estimate the upper bound for the critical time $t_c$, i.e. 
 \be\label{bound}
 t_c  < t(\ell_{{SB}}=1)=\frac{N}{4}\ell_0\sqrt{1-2\ell_0^2}<\frac{N}{4}\ell_0<\frac{N}{4},
 \ee
where the width of the mean-field soliton density corresponding to a symmetry broken state satisfies $\ell_0<1$.

In result it is not possible to observe even a single revolution of the soliton before its decay for any choice of the system parameters. Note that the estimation was made assuming that the soliton structure becomes invisible when its width is equal to the system size. This is why the resulting upper bound for $t_c$ exceeds lifetimes presented in Fig.~\ref{fig_1}. The later were defined through the contrast quantity, Eq.~\eqref{contrast},  which is a more restrictive definition of $t_c$.

To conclude, by means of full many-body simulations we have shown that time crystal dynamics cannot be observed in a soliton model for a finite number of bosons. The reason for that is a quick decay of a spontaneously formed soliton which happens due to quantum many-body effects. We have considered a simple case where the continuous time translation symmetry of the many-body state is spontaneously broken by an initial measurement of a single particle. Although the initial measurement of many particles improves the center of mass localization, it simultaneously decreases the lifetime of the soliton.  Our numerical simulations are consistent with a simple analytical picture for the temporal delocalization of the center of mass of the soliton. In particular,  the numerical results presented here are supported by  the estimation of the upper bound for the lifetime of the soliton. The latter is always smaller than the minimal period of the revolution of the soliton around the ring.

\section*{Acknowledgements}
The work was supported by the National Science Centre, Poland via Projects No.~2018/28/T/ST2/00372 (A.S.) and No.~2018/31/B/ST2/00349
(A.K. and K.S.) and Olle Engkvists stiftelse (A.S.). AK acknowledges the support of Austrian Academy of Sciences’ (ÖAW) ESQ-Discovery Grant.

. 
%%%%%%%%%%%%%%%%%%%%%%%%%%%%%%%%%%%%%%%%%%%%%%%Supported%%%%%%%%%%%%%%%%%%%%%%%%%%%%%%%%%%%%%%%%

%\bibliography{ref_tc_book}

%%%%%%%%%%%%%%%%%%%%%%%%%%%%%%%%%%%%%%%%%%%%%%%%%%%%%%%%%%%%%%%%%%%%%%%%%%%%%%%%%%%%%%%

\end{document}